\documentstyle[12pt]{article}

\newcommand {\nn}    {\nonumber}
\newcommand {\vs}[1]  { \vspace*{#1 cm} }
\newcounter{eq}
\newcounter{sc}

\newcommand {\MPL}  {Mod.Phys.Lett.}
\newcommand {\NP}   {Nucl.Phys.}
\newcommand {\PL}   {Phys.Lett.}
\newcommand {\PR}   {Phys.Rev.}
\newcommand {\PRL}   {Phys.Rev.Lett.}

\def\overleftrightarrow#1{\vbox{\ialign{##\crcr
 $\leftrightarrow$\crcr\noalign{\kern-1pt\nointerlineskip}
 $\hfil\displaystyle{#1}\hfil$\crcr}}}

\newlength{\minitwocolumn}
\begin{document}

%%%%%%%%%%%%%%%%%%%%%%%%%%%%%%%%%%%%%%%%%%%%%%%%%%%%%%%%%%%%%%%%%%
%%%%%%%%%%%%%%%%%%%%%%%% Title %%%%%%%%%%%%%%%%%%%%%%%%%%%%%%%%%%%
%%%%%%%%%%%%%%%%%%%%%%%%%%%%%%%%%%%%%%%%%%%%%%%%%%%%%%%%%%%%%%%%%%

\begin{flushright}
EDO-EP-21\\
September, 1998\\
\end{flushright}
\vspace{30pt}

%\magnification=\magstep1
\pagestyle{empty}
\baselineskip15pt
%\font\cmssB=cmss17
%\font\cmssS=cmss10

\begin{center}
{\large\bf Super D-string Action on $AdS_5 \times S^5$
 \vskip 1mm
}

\vspace{20mm}

Ichiro Oda
          \footnote{
          E-mail address:\ ioda@edogawa-u.ac.jp
                  }
\\
\vspace{10mm}
          Edogawa University,
          474 Komaki, Nagareyama City, Chiba 270-0198, JAPAN \\

\end{center}

%\maketitle

\vspace{15mm}
\begin{abstract}
We present a supersymmetric and $\kappa$-symmetric D-string 
action on $AdS_5 \times S^5$ in supercoset construction. 
As in the previous work of the super D-string action in the
flat background, the super D-string action on $AdS_5 \times S^5$
can be transformed to a form of the IIB Green-Schwarz superstring 
action with the $SL(2,Z)$ covariant tension on $AdS_5 \times S^5$ 
through a duality transformation. In order to understand a part 
of the duality transformation as $SO(2)$ rotation of $N=2$ 
spinor coordinates, it seems to be necessary to fix 
the $\kappa$-symmetry 
in a gauge condition which simplifies the classical action.
This is the article showing for the first time that there exists
S-duality in type IIB superstring theory in a curved background
whose validity has been conjectured in the past but not shown
so far in an explicit way.
\vspace{15mm}

\end{abstract}

\newpage
\pagestyle{plain}
\pagenumbering{arabic}
%\setcounter{page}{1}

%%%%%%%%%%%%%%%%%%%%%%%%%%%%%%%%%%%%%%%%%%%%%%%%%%%%%%%%%%%%%%%%%%
%%%%%%%%%%%%%%%%%%%%%%%% Article %%%%%%%%%%%%%%%%%%%%%%%%%%%%%%%%%
%%%%%%%%%%%%%%%%%%%%%%%%%%%%%%%%%%%%%%%%%%%%%%%%%%%%%%%%%%%%%%%%%%

\rm
%%%%%%%%%%%%%%%%%%%%%%%%%%%%%%%%%%%%%%%%%%%%%%%%%%%%%%%%%%%%%%%%%%%%%
%%%%%%%%%%%%%%%%%%%%%%%%%%%%%%   SEC  1    %%%%%%%%%%%%%%%%%%%%%%%%%%
%%%%%%%%%%%%%%%%%%%%%%%%%%%%%%%%%%%%%%%%%%%%%%%%%%%%%%%%%%%%%%%%%%%%%
\section{Introduction}

Recently an action of type IIB Green-Schwarz superstring \cite{GS} was
constructed in the $AdS_5 \times S^5$ background in terms of
supercoset formalism \cite{Metsaev1}. (See also \cite{Metsaev2}
for super D3-brane action on this background.)
This action has $\kappa$-symmetry as well as two dimensional 
reparametrization invariance as local symmetries, and reduces to 
the conventional type IIB Green-Schwarz superstring action 
in the flat background limit. 
More recently,
the gauge-fixing of $\kappa$-symmetry was developed in two
different approaches, the supersolvable algebra approach \cite{Fre,
Pesando1} and the Killing gauge approach \cite{Kallosh1, Kallosh &
Tseytlin}, 
and afterwards it was shown
that the gauge-fixed actions obtained in the two approaches in fact 
agree through appropriate rearrangement of fields \cite{Pesando2}.

In such a situation
it seems to be timely to construct a super D-string action
in the $AdS_5 \times S^5$ background and then ask ourselves if 
the duality relations such as $SL(2,Z)$ S-duality \cite{Schwarz} 
between the type IIB Green-Schwarz superstring theory and the 
super D-string theory also exist in this curved background as 
in case of the flat background where
supersymmetric and $\kappa$-symmetric D-brane actions 
\cite{Cederwall, Aganagic, Bergshoeff} provided
a good starting point for studying various properties of D-branes
and the web of string dualities. We will show later that this is 
indeed the case.  

The contents of this article are as follows. First of all, 
we construct a super D-string action
on $AdS_5 \times S^5$ explicitly on the basis of the recently
developed supercoset formalism. 
Next we verify that this super D-string action is invariant 
under the $\kappa$-transformation. 
Moreover, we construct the Nambu-Goto form of the type IIB
Green-Schwarz superstring action and its $\kappa$-transformation
in a similar form to the super D-string.
Based on these studies it is shown that the super D-string action
on $AdS_5 \times S^5$ is transformed to the type IIB Green-Schwarz
superstring action with the modified tension on $AdS_5 \times S^5$ 
by performing the duality transformation. In the process, we
need to achieve $SO(2)$ rotation with respect to $N=2$ spinor
coordinates, but to this aim it seems to be necessary to fix 
$\kappa$-symmetry to simplify the classical action.

%%%%%%%%%%%%%%%%%%%%%%%%%%%%%%%%%%%%%%%%%%%%%%%%%%%%%%%%%%%%%%%%%%%%%
%%%%%%%%%%%%%%%%%%%%%%%%%%%%%%   SEC  2    %%%%%%%%%%%%%%%%%%%%%%%%%%
%%%%%%%%%%%%%%%%%%%%%%%%%%%%%%%%%%%%%%%%%%%%%%%%%%%%%%%%%%%%%%%%%%%%%
\section{ Super D-string action }

In this section, we construct the super D-string action in the 
$AdS_5 \times S^5$ background in terms of supercoset formalism
and examine $\kappa$-symmetry of this action.

The $\kappa$-symmetric and reparametrization invariant super D-string 
action in the $AdS_5 \times S^5$ background is given by
%**   2.1 %%%%%%%%%%%%%%%%%%%%%%%%%%%%%%%%%%%%%%%%%%%%%%%%%%%%%%%%%
\begin{eqnarray}
S = S_{DBI} + S_{WZ},
\label{2.1}
\end{eqnarray}
%%%%%%%%%%%%%%%%%%%%%%%%%%%%%%%%%%%%%%%%%%%%%%%%%%%%%%%%%%%%%%%%%%%
with
%**   2.2 %%%%%%%%%%%%%%%%%%%%%%%%%%%%%%%%%%%%%%%%%%%%%%%%%%%%%%%%%
\begin{eqnarray}
&{}& S_{DBI} = -  \int_{M_2} d^2 \sigma 
\sqrt{- \det ( G_{ij} + {\cal F}_{ij} )}, \nn\\
&{}& S_{WZ} =  \int_{M_3} H_3 ({\cal I}) = \int_{M_2 = \partial M_3} 
\Omega_2 ({\cal I}),
\label{2.2}
\end{eqnarray}
%%%%%%%%%%%%%%%%%%%%%%%%%%%%%%%%%%%%%%%%%%%%%%%%%%%%%%%%%%%%%%%%%%%
where $H_3 ({\cal I}) = d \Omega_2 ({\cal I}) = i \bar{L} \wedge \hat{L}
\wedge {\cal I} L$, $i$ and $j$ run over the world-sheet indices
0 and 1, and we have defined 
%**   2.3 %%%%%%%%%%%%%%%%%%%%%%%%%%%%%%%%%%%%%%%%%%%%%%%%%%%%%%%%%
\begin{eqnarray}
{\cal E} = i \sigma_2 = \pmatrix{
0  & 1 \cr -1 & 0 \cr }, \ {\cal I} = \sigma_1 = \pmatrix{
0  & 1 \cr 1 & 0 \cr },  \ {\cal K} = \sigma_3 = \pmatrix{
1  & 0 \cr 0 & -1 \cr }.
\label{2.3}
\end{eqnarray}
%%%%%%%%%%%%%%%%%%%%%%%%%%%%%%%%%%%%%%%%%%%%%%%%%%%%%%%%%%%%%%%%%%%
Here $\sigma_i$ are the Pauli matrices, which operate on $N=2$
spinor indices $I, J = 1, 2$. 
Throughout this article we follow the conventions and notations of 
references \cite{Metsaev1, Metsaev2, Kallosh & Tseytlin}. 
 
In the explicit parametrization $G(X, \Theta) = g(X) e^{\Theta Q}$
where $X^{\hat{m}}$, $\Theta^I$, and $Q_I$ are respectively the 
bosonic and fermionic space-time coordinates and 32-component
supercharges \cite{Metsaev1}, the action (\ref{2.1}) takes the form 
%**   2.4 %%%%%%%%%%%%%%%%%%%%%%%%%%%%%%%%%%%%%%%%%%%%%%%%%%%%%%%%%
\begin{eqnarray}
S = -  \int_{M_2} d^2 \sigma 
\left[ \sqrt{- \det ( G_{ij} + {\cal F}_{ij} )}
-2 i \epsilon^{ij} \int^1_0 ds L^{\hat{a}}_{is}
\bar{\Theta} \Gamma^{\hat{a}} {\cal I} L_{js} \right].
\label{2.4}
\end{eqnarray}
%%%%%%%%%%%%%%%%%%%%%%%%%%%%%%%%%%%%%%%%%%%%%%%%%%%%%%%%%%%%%%%%%%%
The Cartan 1-form superfields $L^I = L^I_{s=1}$ (32-component spinor) 
and $L^{\hat{a}}_{s=1}$ (10 dimensional vector) are given by
\cite{Kallosh2}
%**   2.5 %%%%%%%%%%%%%%%%%%%%%%%%%%%%%%%%%%%%%%%%%%%%%%%%%%%%%%%%%
\begin{eqnarray}
L^I_s &=& \left( \frac{\sinh(s {\cal M})}{\cal M} D\Theta \right)^I, \nn\\
L^{\hat{a}}_s &=& e^{\hat{a}}_{\hat{m}}(X)dX^{\hat{m}}
- 4i {\bar{\Theta}}^I \Gamma^{\hat{a}} \left( \frac{\sinh^2
(\frac{1}{2} s {\cal M})}{{\cal M}^2} D\Theta \right)^I,
\label{2.5}
\end{eqnarray}
%%%%%%%%%%%%%%%%%%%%%%%%%%%%%%%%%%%%%%%%%%%%%%%%%%%%%%%%%%%%%%%%%%%
with 
%**   2.6 %%%%%%%%%%%%%%%%%%%%%%%%%%%%%%%%%%%%%%%%%%%%%%%%%%%%%%%%%
\begin{eqnarray}
({\cal M}^2)^{IL} &=& \epsilon^{IJ} (-\gamma^a \Theta^J
{\bar{\Theta}}^L \gamma^a +  \gamma^{a'} \Theta^J
{\bar{\Theta}}^L \gamma^{a'})
+ \frac{1}{2} \epsilon^{KL} (\gamma^{ab} \Theta^I
{\bar{\Theta}}^K \gamma^{ab} -  \gamma^{a'b'} \Theta^I
{\bar{\Theta}}^K \gamma^{a'b'}), \nn\\
(D\Theta)^I &=& \left[d + \frac{1}{4}(\omega^{ab}\gamma_{ab}
+ \omega^{a'b'}\gamma_{a'b'}) \right] \Theta^I
- \frac{1}{2} i \epsilon^{IJ}(e^a \gamma_a + i e^{a'}
\gamma_{a'}) \Theta^J,
\label{2.6}
\end{eqnarray}
%%%%%%%%%%%%%%%%%%%%%%%%%%%%%%%%%%%%%%%%%%%%%%%%%%%%%%%%%%%%%%%%%%%
where for later convenience we decompose ten dimensional flat 
index $\hat{a}$ in $'5+5'$ and $'4+6'$ ways, $\hat{a} = (a, a')
= (0, \ldots, 4, 5, \ldots, 9) = (p, t) = (0, \ldots, 3, 4, \ldots, 9)$
\cite{Kallosh1, Kallosh & Tseytlin}.
Finally, $G_{ij}$ and ${\cal F}_{01}$ are defined as
%**   2.7 %%%%%%%%%%%%%%%%%%%%%%%%%%%%%%%%%%%%%%%%%%%%%%%%%%%%%%%%%
\begin{eqnarray}
G_{ij} = L^{\hat{a}}_i L^{\hat{a}}_j,  \
{\cal F}_{01} = F_{01} + \epsilon^{ij} \Omega_{ij}({\cal K}),
\label{2.7}
\end{eqnarray}
%%%%%%%%%%%%%%%%%%%%%%%%%%%%%%%%%%%%%%%%%%%%%%%%%%%%%%%%%%%%%%%%%%%
where   
%**   2.8 %%%%%%%%%%%%%%%%%%%%%%%%%%%%%%%%%%%%%%%%%%%%%%%%%%%%%%%%%
\begin{eqnarray}
F_{ij} &=& \partial_i A_j - \partial_j A_i, \nn\\
\Omega_{ij}({\cal K}) &=& i \int^1_0 ds L^{\hat{a}}_{is}
\bar{\Theta} \Gamma^{\hat{a}} {\cal K} L_{js} - 
(i \leftrightarrow j).
\label{2.8}
\end{eqnarray}
%%%%%%%%%%%%%%%%%%%%%%%%%%%%%%%%%%%%%%%%%%%%%%%%%%%%%%%%%%%%%%%%%%%

Actually it is easy to show that the action (\ref{2.4}) precisely 
reduces to the super D-string action on the flat background 
\cite{Cederwall, Aganagic, Bergshoeff} in the flat background limit.
For instance, in that limit the Wess-Zumino term reduces to
%**   2.9 %%%%%%%%%%%%%%%%%%%%%%%%%%%%%%%%%%%%%%%%%%%%%%%%%%%%%%%%%
\begin{eqnarray}
S_{WZ} =  - i \int_{M_2} d^2 \sigma  \epsilon^{ij} \bar{\Theta}
\Gamma^{\hat{a}} {\cal I} \partial_i \Theta \cdot
(\Pi^{\hat{a}}_j + \frac{1}{2} i \bar{\Theta} \Gamma^{\hat{a}} 
\partial_j \Theta),
\label{2.9}
\end{eqnarray}
%%%%%%%%%%%%%%%%%%%%%%%%%%%%%%%%%%%%%%%%%%%%%%%%%%%%%%%%%%%%%%%%%%% 
where $\Pi^{\hat{a}}_i \equiv \partial_i X^{\hat{a}} - i \bar{\Theta} 
\Gamma^{\hat{a}} \partial_i \Theta$, which is exactly the same form
as in the flat background case. This follows from the fact that
in the flat background limit the superfields are given by \cite{Metsaev1}
%**   2.10 %%%%%%%%%%%%%%%%%%%%%%%%%%%%%%%%%%%%%%%%%%%%%%%%%%%%%%%%%
\begin{eqnarray}
L^{\hat{a}}_{is} =  \partial_i X^{\hat{a}} - i s^2 \bar{\Theta}^I 
\Gamma^{\hat{a}} \partial_i \Theta^I, \  L^{I}_{is} = s \partial_i 
\Theta^I.
\label{2.10}
\end{eqnarray}
%%%%%%%%%%%%%%%%%%%%%%%%%%%%%%%%%%%%%%%%%%%%%%%%%%%%%%%%%%%%%%%%%%% 

Now we are ready to present the $\kappa$-transformation of
the super D-string action (\ref{2.4}) whose concrete expressions
are given by
%**   2.11 %%%%%%%%%%%%%%%%%%%%%%%%%%%%%%%%%%%%%%%%%%%%%%%%%%%%%%%%%
\begin{eqnarray}
\delta_{\kappa} \Theta^I = \kappa^I,
\label{2.11}
\end{eqnarray}
%%%%%%%%%%%%%%%%%%%%%%%%%%%%%%%%%%%%%%%%%%%%%%%%%%%%%%%%%%%%%%%%%%%
and the projection $\Gamma$ is
%**   2.12 %%%%%%%%%%%%%%%%%%%%%%%%%%%%%%%%%%%%%%%%%%%%%%%%%%%%%%%%%
\begin{eqnarray}
\Gamma \kappa  = \kappa, \ \Gamma^2 = 1, \ Tr\Gamma = 0, \nn\\
\Gamma = \frac{1}{2} \frac{\epsilon^{ij}}{\sqrt{- \det ( G_{ij} + 
{\cal F}_{ij} )}} (\Gamma_{ij} {\cal I} + {\cal F}_{ij} {\cal E}),
\label{2.12}
\end{eqnarray}
%%%%%%%%%%%%%%%%%%%%%%%%%%%%%%%%%%%%%%%%%%%%%%%%%%%%%%%%%%%%%%%%%%%
with the definition of $\Gamma_{ij} \equiv \frac{1}{2}({\hat{L}}_i
{\hat{L}}_j - {\hat{L}}_j {\hat{L}}_i)$. Moreover, various 
superfields must transform under the $\kappa$-transformation
as follows \cite{Metsaev1, Metsaev2}
%**   2.13 %%%%%%%%%%%%%%%%%%%%%%%%%%%%%%%%%%%%%%%%%%%%%%%%%%%%%%%%%
\begin{eqnarray}
\delta_{\kappa} L^{\hat {a}} &=& 2i \bar{L} \Gamma^{\hat{a}} 
\delta_{\kappa} \Theta, \nn\\
\delta_{\kappa} L &=& d \delta_{\kappa} \Theta - \frac{i}{2} \sigma_+
{\hat{L}} {\cal E} \delta_{\kappa} \Theta + \frac{1}{4}
L^{\hat{a}\hat{b}} \Gamma^{\hat{a}\hat{b}} \delta_{\kappa} \Theta, \nn\\
\delta_{\kappa} \bar{L} &=& d \delta_{\kappa} \bar{\Theta} +
\frac{i}{2} \delta_{\kappa} \bar{\Theta} {\cal E} {\hat{L}}\sigma_+
- \frac{1}{4} \delta_{\kappa} \bar{\Theta} \Gamma^{\hat{a}\hat{b}} 
L^{\hat{a}\hat{b}}, \nn\\
\delta_{\kappa} G_{ij} &=& 2i (\bar{L}_i {\hat{L}}_j + 
\bar{L}_j {\hat{L}}_i) \delta_{\kappa} \Theta, \nn\\
\delta_{\kappa} {\cal F}_{ij} &=& 2i (\bar{L}_i {\cal K} {\hat{L}}_j 
- \bar{L}_j {\cal K} {\hat{L}}_i) \delta_{\kappa} \Theta.
\label{2.13}
\end{eqnarray}
%%%%%%%%%%%%%%%%%%%%%%%%%%%%%%%%%%%%%%%%%%%%%%%%%%%%%%%%%%%%%%%%%%%

Then it is straightforward to show that
%**   2.14 %%%%%%%%%%%%%%%%%%%%%%%%%%%%%%%%%%%%%%%%%%%%%%%%%%%%%%%%%
\begin{eqnarray}
\delta_{\kappa} H_3 ({\cal I}) = d \Lambda_2 ({\cal I}),
\label{2.14}
\end{eqnarray}
%%%%%%%%%%%%%%%%%%%%%%%%%%%%%%%%%%%%%%%%%%%%%%%%%%%%%%%%%%%%%%%%%%%
with $\Lambda_2 ({\cal I}) = 2i \bar{L} \wedge \hat{L} {\cal I}
\delta_{\kappa} \Theta$. Thus the $\kappa$-transformation of the
Wess-Zumino term becomes
%**   2.15 %%%%%%%%%%%%%%%%%%%%%%%%%%%%%%%%%%%%%%%%%%%%%%%%%%%%%%%%%
\begin{eqnarray}
\delta_{\kappa} S_{WZ} = \int_{M_3} d \Lambda_2 ({\cal I})
= 2i \int_{M_2} \bar{L} \wedge \hat{L} {\cal I}
\delta_{\kappa} \Theta.
\label{2.15}
\end{eqnarray}
%%%%%%%%%%%%%%%%%%%%%%%%%%%%%%%%%%%%%%%%%%%%%%%%%%%%%%%%%%%%%%%%%%%
In order to prove the equation (\ref{2.14}), we have made use of
the following Fierz identity for Grassmann odd functions $A, B, C, D$
\cite{Cederwall}
%**   2.16 %%%%%%%%%%%%%%%%%%%%%%%%%%%%%%%%%%%%%%%%%%%%%%%%%%%%%%%%%
\begin{eqnarray}
(\bar{A} \Gamma^{\hat{a}} B) \cdot (\bar{C} \Gamma^{\hat{a}} D)
= -\frac{1}{2} \left[(\bar{A} \Gamma^{\hat{a}} e_I D) \cdot
(\bar{B} \Gamma^{\hat{a}} e_I C) + (\bar{A} \Gamma^{\hat{a}} e_I C) 
\cdot (\bar{B} \Gamma^{\hat{a}} e_I D) \right],
\label{2.16}
\end{eqnarray}
%%%%%%%%%\%%%%%%%%%%%%%%%%%%%%%%%%%%%%%%%%%%%%%%%%%%%%%%%%%%%%%%%%%%
where $e_I = \{1, {\cal E}, {\cal I}, {\cal K}\}$, and 
the Maurer-Cartan equations for $su(2, 2|4)$ superalgebra
\cite{Metsaev1, Metsaev2}
%**   2.17 %%%%%%%%%%%%%%%%%%%%%%%%%%%%%%%%%%%%%%%%%%%%%%%%%%%%%%%%%
\begin{eqnarray}
d L^{\hat {a}} &=& - L^{\hat{a}\hat{b}} \wedge L^{\hat {b}}
- i \bar{L} \Gamma^{\hat{a}} \wedge L, \nn\\ 
d L &=& \frac{i}{2} \sigma_+ {\hat{L}} \wedge {\cal E} L
-\frac{1}{4} L^{\hat{a}\hat{b}} \Gamma^{\hat{a}\hat{b}} \wedge L, \nn\\
d \bar{L} &=& \frac{i}{2} \bar{L} {\cal E} \wedge {\hat{L}} \sigma_+ 
-\frac{1}{4} \bar{L} \Gamma^{\hat{a}\hat{b}} \wedge L^{\hat{a}\hat{b}}.
\label{2.17}
\end{eqnarray}
%%%%%%%%%%%%%%%%%%%%%%%%%%%%%%%%%%%%%%%%%%%%%%%%%%%%%%%%%%%%%%%%%%%
Then it is easy to prove the $\kappa$-invariance of the super
D-string action on $AdS_5 \times S^5$ by showing
%**   2.18 %%%%%%%%%%%%%%%%%%%%%%%%%%%%%%%%%%%%%%%%%%%%%%%%%%%%%%%%%
\begin{eqnarray}
\delta_{\kappa} S_{DBI} + \delta_{\Gamma\kappa} S_{WZ} = 0,
\label{2.18}
\end{eqnarray}
%%%%%%%%%%%%%%%%%%%%%%%%%%%%%%%%%%%%%%%%%%%%%%%%%%%%%%%%%%%%%%%%%%%
where we have made use of a useful identity
%**   2.19 %%%%%%%%%%%%%%%%%%%%%%%%%%%%%%%%%%%%%%%%%%%%%%%%%%%%%%%%%
\begin{eqnarray}
\epsilon^{ij} \epsilon^{kl} = \det G (G^{ik}G^{jl} - 
G^{il}G^{jk}).
\label{2.19}
\end{eqnarray}
%%%%%%%%%%%%%%%%%%%%%%%%%%%%%%%%%%%%%%%%%%%%%%%%%%%%%%%%%%%%%%%%%%%

%%%%%%%%%%%%%%%%%%%%%%%%%%%%%%%%%%%%%%%%%%%%%%%%%%%%%%%%%%%%%%%%%%%%%
%%%%%%%%%%%%%%%%%%%%%%%%%%%%%%   SEC  3    %%%%%%%%%%%%%%%%%%%%%%%%%%
%%%%%%%%%%%%%%%%%%%%%%%%%%%%%%%%%%%%%%%%%%%%%%%%%%%%%%%%%%%%%%%%%%%%%
\section{ The Nambu-Goto form of type IIB superstring}

In the pioneering paper \cite{Metsaev1}, 
the Polyakov form of type IIB Green-Schwarz 
superstring action was constructed in the $AdS_5 \times S^5$ background 
in terms of supercoset formalism. As a trivial extension of it, we shall
present the Nambu-Goto form and construct its local $\kappa$-transformation
which is of the form similar to that of super D-string made in the
previous section.

The type IIB Green-Schwarz superstring action on $AdS_5 \times S^5$ 
is given by in the Polyakov form \cite{Metsaev1}
%**   3.1 %%%%%%%%%%%%%%%%%%%%%%%%%%%%%%%%%%%%%%%%%%%%%%%%%%%%%%%%%
\begin{eqnarray}
S = S_{Poly} + S_{WZ},
\label{3.1}
\end{eqnarray}
%%%%%%%%%%%%%%%%%%%%%%%%%%%%%%%%%%%%%%%%%%%%%%%%%%%%%%%%%%%%%%%%%%%
with
%**   3.2 %%%%%%%%%%%%%%%%%%%%%%%%%%%%%%%%%%%%%%%%%%%%%%%%%%%%%%%%%
\begin{eqnarray}
&{}& S_{Poly} = - \frac{1}{2} \int_{M_2} d^2 \sigma 
\sqrt{- g} g^{ij} L^{\hat{a}}_i L^{\hat{a}}_j, \nn\\
&{}& S_{WZ} =  \int_{M_3} H_3 (-{\cal K}) = \int_{M_2 = \partial M_3} 
\Omega_2 (-{\cal K}),
\label{3.2}
\end{eqnarray}
%%%%%%%%%%%%%%%%%%%%%%%%%%%%%%%%%%%%%%%%%%%%%%%%%%%%%%%%%%%%%%%%%%%
where $H_3(-{\cal K})$ and $\Omega_2 (-{\cal K})$ are defined as 
in the previous section but with the different argument 
$-{\cal K}$. 
Usually the local $\kappa$-transformation is defined by \cite{Metsaev1}
%**   3.3 %%%%%%%%%%%%%%%%%%%%%%%%%%%%%%%%%%%%%%%%%%%%%%%%%%%%%%%%%
\begin{eqnarray}
\delta_{\kappa} \Theta^I &=& 2 {\hat{L}}_i \kappa^{iI}, \nn\\
\delta_{\kappa}(\sqrt{- g} g^{ij}) &=& -16 i \sqrt{- g} 
(P^{jk}_{-} \bar{L}^1_k \kappa^{i1} + P^{jk}_{+} \bar{L}^2_k 
\kappa^{i2}), 
\label{3.3}
\end{eqnarray}
%%%%%%%%%%%%%%%%%%%%%%%%%%%%%%%%%%%%%%%%%%%%%%%%%%%%%%%%%%%%%%%%%%%
where
%**   3.4 %%%%%%%%%%%%%%%%%%%%%%%%%%%%%%%%%%%%%%%%%%%%%%%%%%%%%%%%%
\begin{eqnarray}
P^{ij}_{-} \kappa^1_j &=& \kappa^{i1}, \nn\\
P^{ij}_{+} \kappa^2_j &=& \kappa^{i2},  
\label{3.4}
\end{eqnarray}
%%%%%%%%%%%%%%%%%%%%%%%%%%%%%%%%%%%%%%%%%%%%%%%%%%%%%%%%%%%%%%%%%%%
with the definition of the projection operator
$P^{ij}_{\pm} \equiv \frac{1}{2}(g^{ij} \pm \frac{1}{\sqrt{-g}}
\epsilon^{ij})$. Of course, the $\kappa$-transformation for the
superfields is defined as in (\ref{2.13}).

Solving the field equation with respect to the auxiliary world-sheet
 metric $g_{ij}$ leads to the Nambu-Goto form of the type IIB
Green-Schwarz superstring action
%**   3.5 %%%%%%%%%%%%%%%%%%%%%%%%%%%%%%%%%%%%%%%%%%%%%%%%%%%%%%%%%
\begin{eqnarray}
S &=& S_{NG} + S_{WZ}  \nn\\
  &=& - \int_{M_2} d^2 \sigma \sqrt{- \det G_{ij}} + S_{WZ},
\label{3.5}
\end{eqnarray}
%%%%%%%%%%%%%%%%%%%%%%%%%%%%%%%%%%%%%%%%%%%%%%%%%%%%%%%%%%%%%%%%%%%
where $G_{ij} = L^{\hat{a}}_i L^{\hat{a}}_j$ and the Wess-Zumino
term remains unchanged. 

It is interesting to notice that this action (\ref{3.5}) is
invariant under a similar $\kappa$-transformation to as
in the super D-string action (\ref{2.11}),(\ref{2.12}). 
Namely, the local $\kappa$-transformation can now be described by
%**   3.6 %%%%%%%%%%%%%%%%%%%%%%%%%%%%%%%%%%%%%%%%%%%%%%%%%%%%%%%%%
\begin{eqnarray}
\delta_{\kappa} \Theta^I = \kappa^I,
\label{3.6}
\end{eqnarray}
%%%%%%%%%%%%%%%%%%%%%%%%%%%%%%%%%%%%%%%%%%%%%%%%%%%%%%%%%%%%%%%%%%%
and the projection $\Gamma$ is defined as 
%**   3.7 %%%%%%%%%%%%%%%%%%%%%%%%%%%%%%%%%%%%%%%%%%%%%%%%%%%%%%%%%
\begin{eqnarray}
\Gamma \kappa  = \kappa, \ \Gamma^2 = 1, \ Tr\Gamma = 0, \nn\\
\Gamma = -\frac{1}{2} \frac{\epsilon^{ij}}{\sqrt{- \det G_{ij}}} 
\Gamma_{ij} {\cal K}.
\label{3.7}
\end{eqnarray}
%%%%%%%%%%%%%%%%%%%%%%%%%%%%%%%%%%%%%%%%%%%%%%%%%%%%%%%%%%%%%%%%%%%
It is straightforward to show that $\delta_{\kappa} {\cal L}_{WZ}
= -2i \epsilon^{ij} \bar{L}_i {\hat{L}_j} {\cal{K}} \delta_{\kappa} 
\Theta$ which is precisely canceled against a contribution from
the $\kappa$ variation of the Nambu-Goto action.

%%%%%%%%%%%%%%%%%%%%%%%%%%%%%%%%%%%%%%%%%%%%%%%%%%%%%%%%%%%%%%%%%%%%%
%%%%%%%%%%%%%%%%%%%%%%%%%%%%%%   SEC  4    %%%%%%%%%%%%%%%%%%%%%%%%%%
%%%%%%%%%%%%%%%%%%%%%%%%%%%%%%%%%%%%%%%%%%%%%%%%%%%%%%%%%%%%%%%%%%%%%
\section{ Duality transformation between super D-string and type
IIB Green-Schwarz superstring actions}

In the previous sections, we have investigated super D-string
and type IIB Green-Schwarz superstring actions in the $AdS_5 \times 
S^5$ background. In this section, we wish to clarify the relationship
between the two actions, in other words, the duality transformation.
For most of works done so far, analysis of the duality transformation
properties of super Dp-branes has been classical and limited to
the flat background although the results are expected not to depend
on these restrictions \cite{Aganagic}. 
Our purpose in this section is to remove
these restrictions and carry out the analysis not only in a quantum
mechanical way but also in a curved background.
Of course our analysis is still unsatisfactory in that we deal
with only the specific background $AdS_5 \times S^5$ and super 
D1-brane, but it would be the first important step towards 
the full analysis of the duality transformation properties of 
super Dp-branes in general background.

Now we are in a position to show how the super D-string action 
(\ref{2.4}) becomes a fundamental superstring action (\ref{3.5}) 
with the $SL(2, Z)$ covariant tension by using the path integral. 
We shall use the first-order Hamiltonian formalism
of the path integral which was found by de Alwis and Sato 
\cite{de Alwis} in the bosonic case and later applied to the
supersymmetric case by the present author \cite{Oda1}.
Note that this formalism does not rely on any approximation
at least in case of string even if it is necessary to use a 
saddle point approximation when we want to apply this method to 
super Dp-branes with $p > 1$ because of the nonlinear feature 
of the p-brane actions.

Here it is worthwhile to comment on why the first-order Hamiltonian
formalism on the $U(1)$ gauge sector plays an 
important role in the analysis of the duality transformation 
of D-branes. One reason comes from the fact that 
the duality is a 'symmetry' not in the Lagrangian but in the 
Hamiltonian.
In other words, the duality is a 'symmetry' holding only at the
level of classical field equations.
The other reason is that the major difference between super 
D-branes and super F-branes exists in the presence of $U(1)$ 
gauge field in the former. 
Hence in order to understand the duality transformation between 
the two actions it is enough to use the first-order Hamiltonian
formalism only on the $U(1)$ gauge sector in the
super D-branes. 

According to the Hamiltonian formalism, let us start by introducing
the canonical conjugate momenta $\pi^i$ corresponding to the 
gauge field $A_i$ defined as
%**   4.1 %%%%%%%%%%%%%%%%%%%%%%%%%%%%%%%%%%%%%%%%%%%%%%%%%%%%%%%%%
\begin{eqnarray}
\pi^i \equiv \frac{\partial S^{D-string}}
{\partial \dot{A}_i} =  \frac{\partial S_{DBI}}
{\partial \dot{A}_i},
\label{4.1}
\end{eqnarray}
%%%%%%%%%%%%%%%%%%%%%%%%%%%%%%%%%%%%%%%%%%%%%%%%%%%%%%%%%%%%%%%%%%%
where we used the fact that the Wess-Zumino term is independent
of the gauge potential. Note that compared to the flat background
the position of index $i$ is important in the curved one.
Then the canonical conjugate momenta $\pi^i$ are calculated to
be
%**   4.2 %%%%%%%%%%%%%%%%%%%%%%%%%%%%%%%%%%%%%%%%%%%%%%%%%%%%%%%%%
\begin{eqnarray}
\pi^0 = 0, \ \pi^1 = \frac{{\cal F}_{01}}
{\sqrt{-\det ( G_{ij} + {\cal F}_{ij} )}},
\label{4.2}
\end{eqnarray}
%%%%%%%%%%%%%%%%%%%%%%%%%%%%%%%%%%%%%%%%%%%%%%%%%%%%%%%%%%%%%%%%%%%
from which the Hamiltonian has the form 
%**   4.3 %%%%%%%%%%%%%%%%%%%%%%%%%%%%%%%%%%%%%%%%%%%%%%%%%%%%%%%%%
\begin{eqnarray}
{\cal H} = \sqrt{1 + (\pi^1)^2} \sqrt{- \det G_{ij}} 
- \epsilon^{ij} \Omega_{ij} (\pi^1 {\cal K} + {\cal I}) 
- A_0 \partial_1 \pi^1 + \partial_1 (A_0 \pi^1),
\label{4.3}
\end{eqnarray}
%%%%%%%%%%%%%%%%%%%%%%%%%%%%%%%%%%%%%%%%%%%%%%%%%%%%%%%%%%%%%%%%%%%
where we have chosen the positive sign in front of the first
term without loss of generality since this sign ambiguity is 
related to overall normalization of the action.
 
At this stage, the partition function is defined by the first-order 
Hamiltonian form with respect to only the gauge field as follows:
%**   4.4 %%%%%%%%%%%%%%%%%%%%%%%%%%%%%%%%%%%%%%%%%%%%%%%%%%%%%%%%%
\begin{eqnarray}
Z &=& \frac{1}{\int {\cal D}\pi^0} \int {\cal D}\pi^0
{\cal D}\pi^1 {\cal D}A_0 {\cal D}A_1 
\exp{ i \int d^2 \sigma ( \pi^1 \partial_0 A_1 - {\cal H} ) } \nn\\
&=& \int {\cal D}\pi^1 {\cal D}A_0 {\cal D}A_1 
\exp{ i \int d^2 \sigma} \nn\\
& & {} \left[ - A_1 \partial_0 \pi^1 + A_0 \partial_1
\pi^1 -  \sqrt{1 + (\pi^1)^2} \sqrt{- \det G_{ij}} 
+ \epsilon^{ij} \Omega_{ij} (\pi^1 {\cal K} + {\cal I})
- \partial_1 ( A_0 \pi^1 ) \right] ,
\label{4.4}
\end{eqnarray}
%%%%%%%%%%%%%%%%%%%%%%%%%%%%%%%%%%%%%%%%%%%%%%%%%%%%%%%%%%%%%%%%%%%
where we have canceled the trivial gauge group volume. 
Note that if we take the
boundary conditions for $A_0$ such that 
the last surface term in the exponential identically vanishes, 
then we can carry out the integrations over $A_i$ explicitly, 
which gives rise to $\delta$ functions
%**   4.5 %%%%%%%%%%%%%%%%%%%%%%%%%%%%%%%%%%%%%%%%%%%%%%%%%%%%%%%%%
\begin{eqnarray}
Z &=& \int {\cal D}\pi^1 \delta(\partial_0 \pi^1) 
\delta(\partial_1 \pi^1) \exp{ i \int d^2 \sigma 
\left[ -\sqrt{1 + (\pi^1)^2} \sqrt{- \det G_{ij}} 
+ \epsilon^{ij} \Omega_{ij} (\pi^1 {\cal K} + {\cal I}) \right] }.
\label{4.5}
\end{eqnarray}
%%%%%%%%%%%%%%%%%%%%%%%%%%%%%%%%%%%%%%%%%%%%%%%%%%%%%%%%%%%%%%%%%%%
Note that the existence of the $\delta$ functions reduces the integral 
over $\pi^1$ to the one over only its zero-modes. If we require
that one space component is compactified on a circle, these
zero-modes are quantized to be integers \cite{Witten}.
Consequently, the partition function becomes
%**   4.6 %%%%%%%%%%%%%%%%%%%%%%%%%%%%%%%%%%%%%%%%%%%%%%%%%%%%%%%%%
\begin{eqnarray}
Z &=& \displaystyle{ \sum_{m \in {\bf Z}} } \exp{ i \int d^2 \sigma 
\left[ -\sqrt{1 + m^2} \sqrt{- \det G_{ij}} 
+ \epsilon^{ij} \Omega_{ij} (m {\cal K} + {\cal I}) \right] }.
\label{4.6}
\end{eqnarray}
%%%%%%%%%%%%%%%%%%%%%%%%%%%%%%%%%%%%%%%%%%%%%%%%%%%%%%%%%%%%%%%%%%%
Since the eigenvalues of $m {\cal K} + {\cal I}$ are
$\pm \sqrt{1 + m^2} {\cal K}$, we can redefine
%**   4.7 %%%%%%%%%%%%%%%%%%%%%%%%%%%%%%%%%%%%%%%%%%%%%%%%%%%%%%%%%
\begin{eqnarray}
m {\cal K} + {\cal I} \equiv -\sqrt{1 + m^2} {\cal K}.
\label{4.7}
\end{eqnarray}
%%%%%%%%%%%%%%%%%%%%%%%%%%%%%%%%%%%%%%%%%%%%%%%%%%%%%%%%%%%%%%%%%%%
Then we finally arrive at the patition function
%**   4.8 %%%%%%%%%%%%%%%%%%%%%%%%%%%%%%%%%%%%%%%%%%%%%%%%%%%%%%%%%
\begin{eqnarray}
Z &=& \displaystyle{ \sum_{m \in {\bf Z}} } \exp{ i \int d^2 \sigma 
\sqrt{1 + m^2} \left( - \sqrt{- \det G_{ij}} 
- \epsilon^{ij} \Omega_{ij}({\cal K}) \right) }.
\label{4.8}
\end{eqnarray}
%%%%%%%%%%%%%%%%%%%%%%%%%%%%%%%%%%%%%%%%%%%%%%%%%%%%%%%%%%%%%%%%%%%
{}From this expression of the partition function, we can read off
the effective action
%**   4.9 %%%%%%%%%%%%%%%%%%%%%%%%%%%%%%%%%%%%%%%%%%%%%%%%%%%%%%%%%
\begin{eqnarray}
S = - \sqrt{1 + m^2} \int d^2 \sigma \left(  \sqrt{- \det 
G_{ij}} + \epsilon^{ij} \Omega_{ij}({\cal K}) \right).
\label{4.9}
\end{eqnarray}
%%%%%%%%%%%%%%%%%%%%%%%%%%%%%%%%%%%%%%%%%%%%%%%%%%%%%%%%%%%%%%%%%%%
This is nothing but type IIB Green-Schwarz superstring action
(\ref{3.5})
with the modified tension $\sqrt{1 + m^2}$.  This agrees with
the tension formula for the $SL(2,Z)$ S-duality spectrum
of strings  in the flat background \cite{Schwarz} provided 
that we identify the
integer value $\pi^1 = m$ as corresponding to the $(m,1)$
string. To show clearly that the tension
obtained at hand is the $SL(2,Z)$ covariant tension, it
would be more convenient to start with the following classical action
instead of the action (\ref{2.4})
%**   4.10 %%%%%%%%%%%%%%%%%%%%%%%%%%%%%%%%%%%%%%%%%%%%%%%%%%%%%%%%%
\begin{eqnarray}
S = -  n \int_{M_2} d^2 \sigma 
\left[ e^{-\phi} \bigl( \sqrt{- \det ( G_{ij} + {\cal F}_{ij} )}
-2 i \epsilon^{ij} \int^1_0 ds L^{\hat{a}}_{is}
\bar{\Theta} \Gamma^{\hat{a}} {\cal I} L_{js} \bigr)
+ \frac{1}{2} \epsilon^{ij} \chi F_{ij} \right],
\label{4.10}
\end{eqnarray}
%%%%%%%%%%%%%%%%%%%%%%%%%%%%%%%%%%%%%%%%%%%%%%%%%%%%%%%%%%%%%%%%%%%
where $n$ is an integer, and we confined ourselves to be
only a constant dilaton $\phi$ and a constant axion $\chi$
such that the action (\ref{4.10}) is still invariant under
the same $\kappa$-transformation as before. 
Then following the same path
of thoughts as above, we can obtain the manifestly $SL(2,Z)$
covariant tension
%**   4.11 %%%%%%%%%%%%%%%%%%%%%%%%%%%%%%%%%%%%%%%%%%%%%%%%%%%%%%%%%
\begin{eqnarray}
T = \sqrt{ (m + n \chi)^2 +  n^2 e^{-2\phi}}.
\label{4.11}
\end{eqnarray}
%%%%%%%%%%%%%%%%%%%%%%%%%%%%%%%%%%%%%%%%%%%%%%%%%%%%%%%%%%%%%%%%%%% 

Here we would like to emphasize two important points. One point is 
that we have shown that there exists $SL(2,Z)$ S-duality in type
IIB superstring even in the $AdS_5 \times S^5$ background. We think
that this statement is quite nontrivial and important for future 
development of string theory in curved background geometry. 
Another important point is 
that we have obtained this result without appealing to any 
approximation so that the effective action (\ref{4.9}) is quantum 
mechanically equivalent to the super D-string action (\ref{2.4}).

It may appear that we have succeeded in deriving the $SL(2,Z)$
S duality of type IIB superstring theory at least within the 
present context. However, a reality is not simple. That is,
in the above procedure of derivation we have assumed one 
dubious step tacitly, which amounts to the step of the redefinition  
(\ref{4.7}). Note that to achieve this redefinition successfully
we have to change the basis of spinor variables by an appropriate 
orthogonal matrix.
In case of the flat background this redefinition 
is easily carried out in terms of an $SO(2)$ rotation of the 
spinor coordinates $\Theta^I$ because in this case the Wess-Zumino
term consists of a function of the simple forms like
$\bar{\Theta}^I \Theta^I$ and $\bar{\Theta}^I {\cal I}^{IJ}
\Theta^J$ as in Eq.(\ref{2.9}) \cite{Oda1}. 
In contrast, in case of $AdS_5 \times S^5$
background, as seen in Eqs.(\ref{2.5}),(\ref{2.6}), the spinor and vector
superfields involved in the Wess-Zumino term have quite complicated
dependence on the spinor variables $\Theta^I$ so that 
although it may not be impossible it seems to be difficult
to perform the $SO(2)$ rotation in the classical action. 

Then how do we improve this situation? One idea is to fix the 
$\kappa$-symmetry to make the classical action 
of the super D-string simpler,
and then carry out the $SO(2)$ rotation. Fortunately, a consistent
quantization procedure has recently appeared to 
the type IIB Green-Schwarz superstring action 
on $AdS_5 \times S^5$ \cite{Kallosh1, Kallosh & Tseytlin}. 
This quantization method
can also be taken over the case of the super D-string on
$AdS_5 \times S^5$ in a direct manner. 

Let us explain how to carry out the $SO(2)$ rotation in the process
of the gauge-fixing procedure of the $\kappa$-symmetry.
Before doing it, let us consider what othogonal matrix makes 
$m {\cal K} + {\cal I}$ an othogonal form. By solving the
eigenvalue equation, it is easy to show that 
$U = \frac{1}{\sqrt{1 + (m - \sqrt{1+m^2})^2}}[(m - \sqrt{1+m^2})1
- {\cal E}]$ works well, that is,
%**   4.12 %%%%%%%%%%%%%%%%%%%%%%%%%%%%%%%%%%%%%%%%%%%%%%%%%%%%%%%%%
\begin{eqnarray}
U^T (m {\cal K} + {\cal I}) U = - \sqrt{1 + m^2} {\cal K}.
\label{4.12}
\end{eqnarray}
%%%%%%%%%%%%%%%%%%%%%%%%%%%%%%%%%%%%%%%%%%%%%%%%%%%%%%%%%%%%%%%%%%%

Next let us recall how the $\kappa$-symmetry is fixed using 
the 'parallel to D3-brane' $\Gamma$-matrix projector ${\cal P}^{IJ}_{\pm}$
\cite{Kallosh1, Kallosh & Tseytlin}.
The gauge condition of the $\kappa$-symmetry is chosen to be 
%**   4.13 %%%%%%%%%%%%%%%%%%%%%%%%%%%%%%%%%%%%%%%%%%%%%%%%%%%%%%%%%
\begin{eqnarray}
\Theta^I_{-} = 0,
\label{4.13}
\end{eqnarray}
%%%%%%%%%%%%%%%%%%%%%%%%%%%%%%%%%%%%%%%%%%%%%%%%%%%%%%%%%%%%%%%%%%%
where 
%**   4.14 %%%%%%%%%%%%%%%%%%%%%%%%%%%%%%%%%%%%%%%%%%%%%%%%%%%%%%%%%
\begin{eqnarray}
\Theta^I_{\pm} \equiv  {\cal P}^{IJ}_{\pm} \Theta^J, \ 
{\cal P}^{IJ}_{\pm} \equiv \frac{1}{2}(\delta^{IJ}
\pm \Gamma_{0123} \epsilon^{IJ}).
\label{4.14}
\end{eqnarray}
%%%%%%%%%%%%%%%%%%%%%%%%%%%%%%%%%%%%%%%%%%%%%%%%%%%%%%%%%%%%%%%%%%%
Then, it can be shown that the Wess-Zumino term becomes to
be a quadratic form with respect to $\theta^I_{+}$ by the
change of the fermionic variables from $\Theta^I$ to $\theta^I$ 
where $\Theta^I_{+} = y^{\frac{1}{2}} \theta^I_{+}$ 
and at the same time using the relations 
%**   4.15 %%%%%%%%%%%%%%%%%%%%%%%%%%%%%%%%%%%%%%%%%%%%%%%%%%%%%%%%%
\begin{eqnarray} 
L^t_{is} = \frac{1}{y} \partial_i y^t, \  L^I_{js} = s y^{\frac{1}{2}}
\partial_i \theta^I_{+},
\label{4.15}
\end{eqnarray}
%%%%%%%%%%%%%%%%%%%%%%%%%%%%%%%%%%%%%%%%%%%%%%%%%%%%%%%%%%%%%%%%%%%
which hold in the gauge choice (\ref{4.13}). 
(See the original references \cite{Kallosh1, Kallosh & Tseytlin}
for more detail.)  
The important observation here is that the projector 
${\cal P}^{IJ}_{\pm}$ is invariant under the orthogonal
transformation $U$, i.e., $U^T {\cal P}_{\pm} U = {\cal P}_{\pm}$.
This fact implies that if we change the spinor coordinates
by $\Theta^I = U^{IJ} \tilde{\Theta}^J$, 
we have the relation $\Theta^I_{\pm} = U^{IJ} \tilde{\Theta}^J_{\pm}$.
As a result, in the gauge condition (\ref{4.13}), we reach the
important relation $\theta^I_{+} = U^{IJ} \tilde{\theta}^J_{+}$.
By using this relation as well as other ones, we can prove that
the replacement (\ref{4.7}) is indeed 
legitimated in terms of the $SO(2)$ rotation of the
spinor coordinates. (Incidentally, the Nambu-Goto action is
invariant under this $SO(2)$ rotation as can be checked easily.)
Since the above arguments are a little formal,
let us expose the related equations in order according to the
above arguments in what follows:
%**   4.16 %%%%%%%%%%%%%%%%%%%%%%%%%%%%%%%%%%%%%%%%%%%%%%%%%%%%%%%%%
\begin{eqnarray}
I &\equiv& \epsilon^{ij} \Omega_{ij}(m{\cal K}+{\cal I}) \nn\\
&=& 2i \int^1_0 ds \epsilon^{ij} L^{\hat{a}}_{is} \bar{\Theta} 
\Gamma^{\hat{a}} (m{\cal K}+{\cal I}) L_{js} \nn\\
&=& i \epsilon^{ij} \partial_i y^t \bar{\theta}_{+} \Gamma^t 
(m{\cal K}+{\cal I}) \partial_j \theta_{+} \nn\\
&=& i \epsilon^{ij} \partial_i y^t \bar{\tilde{\theta}}_{+}
\Gamma^t (- \sqrt{1 + m^2}) {\cal K} \partial_j 
\tilde{\theta}_{+}.
\label{4.16}
\end{eqnarray}
%%%%%%%%%%%%%%%%%%%%%%%%%%%%%%%%%%%%%%%%%%%%%%%%%%%%%%%%%%%%%%%%%%%

Finally, we wish to close this section by commenting on
one problem. In the previous work \cite{Oda1}, it was
shown that in the flat background geometry
the super D-string action is exactly equivalent
to the type IIB Green-Schwarz superstring action with
some "theta term" in terms of the same path integral method.
It is then natural to ask what happens to the case of 
$AdS_5 \times S^5$ background. 
We can easily show that this is not always the
case. In fact, to demonstrate the equivalence one needs to
perform a scale transformation, but the nonlinear sigma 
action (or the Nambu-Goto action) in the $AdS_5 \times S^5$ 
background is scale invariant
so that it is impossible to make the super D-string action
coincide with the type IIB Green-Schwarz superstring action
with "theta term" in the case of the $AdS_5 \times S^5$ 
background.

%%%%%%%%%%%%%%%%%%%%%%%%%%%%%%%%%%%%%%%%%%%%%%%%%%%%%%%%%%%%%%%%%%%%%
%%%%%%%%%%%%%%%%%%%%%%%%%%%%%%   SEC  5    %%%%%%%%%%%%%%%%%%%%%%%%%%
%%%%%%%%%%%%%%%%%%%%%%%%%%%%%%%%%%%%%%%%%%%%%%%%%%%%%%%%%%%%%%%%%%%%%
\section{ Discussions }

In this paper, we have constructed a supersymmetric and 
$\kappa$-symmetric D-string action in the $AdS_5 \times S^5$ 
background in supercoset construction. Starting with the
super D-string action it has been shown that one can obtain
the $SL(2,Z)$ multiplet of type IIB strings with the correct
tensions by performing the duality transformation. One of the
most appealing points in this paper is that we have shown the 
existence of the $SL(2,Z)$ multiplet of type IIB strings even 
in the $AdS_5 \times S^5$ background, which was already expected
in the past but not verified explicitly \cite{Aganagic}.
We believe that we have shed some light on the S-duality 
transformation between the super D-string and fundamental 
Green-Schwarz superstring in a nontrivial curved
background.

Finally we would like to make comments on future works.
One interesting direction is to construct general super
D-brane actions on the $AdS_5 \times S^5$ and investigate
various duality transformations. In paticular, it is
expected that the super D3-brane action \cite{Metsaev2}
transforms in the same way as in the super D-string.
Moreover, the super D2-brane and D4-brane actions would
transform in a manner that is expected from the relation
between type IIA superstring theory and 11 dimensional 
M theory. This work is now under active investigation and 
will be reported in a separate publication \cite{Oda2}. 

Another valuable future work is to perform the quantization
of general super D-brane actions like the super D-string
and type IIB Green-Schwarz superstring on $AdS_5 \times 
S^5$. We usually think that these p-brane actions with $p>1$ are 
unrenormalizable so that new dynamical degrees of freedom
appear in the short distance region, and consequently 
problem of quantization is physically uninteresting. 
But it is worthwhile to point out that these arguments 
are entirely based on discussions of the field 
theory in the flat background, so the quantization
of higher (super)p-branes on a curved space-time maniflolds
deserves further studies in future.
For instance, it seems to be difficult to quantize the 
M5-brane action \cite{Mario} in a Lorentz covariant manner,
but it may be possible to perform the quantization if
one couples supergravity background to the theory.
Actually, in the quantization
method adopted in \cite{Kallosh & Tseytlin} the Killing
symmetry of the background metric field plays an
essential role.   

In addition to these future works, we wish to utilize the results
obtained in curved background in order to understand 
background independent matrix models \cite{Oda3}
in a more complete way. 
The progress is still under way but without fruits at present.  

\vs 1
%%%%%%%%%%%%%%%%%%%%%%%%%%%%%%%%%%%%%%%%%%%%%%%%%%%%%%%%%%%%%%%%%%
%%%%%%%%%%%%%%%%%%%%%%%% Acknowledgement %%%%%%%%%%%%%%%%%%%%%%%%%%%%%
%%%%%%%%%%%%%%%%%%%%%%%%%%%%%%%%%%%%%%%%%%%%%%%%%%%%%%%%%%%%%%%%%%
\begin{flushleft}
{\bf Acknowledgement}
\end{flushleft}
We are indebted to M. Tonin for valuable discussions 
and kind hospitality at Padova University where a part of
this work has been done. 
This work was supported in part by Grant-Aid for Scientific 
Research from Ministry of Education, Science and Culture 
No.09740212 and collaboration research grant from Edogawa
University.

\vs 1
%%%%%%%%%%%%%%%%%%%%%%%%%%%%%%%%%%%%%%%%%%%%%%%%%%%%%%%%%%%%%%%%%%
%%%%%%%%%%%%%%%%%%%%%%%% reference %%%%%%%%%%%%%%%%%%%%%%%%%%%%%%%
%%%%%%%%%%%%%%%%%%%%%%%%%%%%%%%%%%%%%%%%%%%%%%%%%%%%%%%%%%%%%%%%%%

\end{document}